\documentclass{pointing}

\usepackage{epsfig}

\begin{document}

\begin{frontmatter}

\title{Locating TeV $\gamma$-ray Sources with\\
Sub-Arcminute Precision:\\
the Pointing Calibration of the HEGRA System of\\
Imaging Atmospheric Cherenkov Telescopes}

\author{G. P\"uhlhofer},
\author{A. Daum},
\author{G. Hermann},
\author{M. He\ss},
\author{W. Hofmann},
\author{C. K\"ohler},
\author{M. Panter}

\address{Max-Planck-Institut f\"ur Kernphysik, P.O. Box 103980,
        D-69029 Heidelberg, Germany}

\begin{abstract}
Stereoscopic viewing of TeV $\gamma$-ray air showers
with systems of Imaging Atmospheric Cherenkov Telescopes (IACTs)
allows to reconstruct the origin of individual primary particles
with an accuracy of $0.1^{\circ}$ or better. The 
shower impact point can be determined within 15 meters.
To actually achieve this resolution, the pointing of the
telescopes of an IACT system needs to be controlled with 
high precision. For the HEGRA IACT system, a procedure to calibrate 
telescope pointing was established, using bright stars
distributed over the sky as references. On the basis of
these measurements, one determines parameters of a
correction function which is valid for the complete hemisphere.
After correction a pointing accuracy of $0.01^{\circ}$ is achieved.
\end{abstract}

\end{frontmatter}

\section{Introduction}
\label{intro}
The imaging Cherenkov technique has proven the most powerful
tool for the detection of air showers induced by TeV $\gamma$-rays.
Several sources have been detected in the last few years~\cite{IACTs}.

A Cherenkov telescope uses a (usually tesselated) 
mirror to project the
Cherenkov light emitted by the air shower particles
into a camera. The camera consists of a matrix of photomultipliers (PM)
which record the Cherenkov light flash, efficiently rejecting night-sky 
background photons due to the short gating times. The image represents a
2-dimensional projection of the shower. The image shape is used
to discriminate between $\gamma$-ray showers and hadron induced showers which
cause a background signal of 99.9\% of the total rate. The position
of the image in the camera determines an area on the sky where
the primary came from.

The most common analysis technique for single-telescope data 
(see, e.g., \cite{Hillas,Punch,Petry})
is mainly applicable to the search for point sources. 
The $\alpha$ parameter - the angle between the vector from the image
center of gravity to the image of a candidate point source, and the
image axis -
is used to select events pointing to the source image.
In such analyses, telescope mispointing can be
tolerated up to $\approx0.1^{\circ}$, without drastic loss in performance~\cite{Petry}.

Stereoscopic viewing of showers 
with two or more telescopes yields far more detailed
and unambiguous information about each TeV event. It allows the determination
of the origin of the primary $\gamma$ particle with an accuracy of better
than $0.1^{\circ}$. The shower core which is defined as the virtual impact
point of the primary onto the ground can be reconstructed within 15 meters.
Besides that one gains very precise information about the energy of the primary
and about its nature.
By averaging over many events, source locations can be
determined with subarcminute accuracy~\cite{Akerlof,Crab_system,Mkn501_system}.

The HEGRA\footnote{HEGRA stands for High Energy Gamma Ray Astronomy, a collaboration of
Max Planck Institutes for Nuclear Physics in Heidelberg and for 
Physics in Munich, Universities of Kiel, Hamburg and Madrid,
BUGH Wuppertal, and Yerevan Physics Institute}
collaboration has built a system of six Imaging Atmospheric Cherenkov
Telescopes~\cite{CT_system}, of which up to now four
telescopes run in coincident
stereoscopic mode; the other two are operated independently. 
The following paper deals with the pointing calibration 
procedure which had to be developed in order to reach the intrinsic
resolution of the stereoscopic system.

Section 2 briefly describes the mounts of the HEGRA IACTS and the 
telescope tracking philosophy.
In section 3 the procedure for pointing calibration is described.
Section 4 presents the model which is used to parametrise the
telescopes' mispointing. In Section 5 the results of the calibrations
made so far are discussed, especially concerning
the mechanical performance of the telescope mounts. Section 6
demonstrates the quality of the pointing calibration by its application
to data of two TeV $\gamma$-ray sources, the Crab pulsar and the active 
galaxy Mkn 501.

\section{Telescope mounts and tracking}
\label{concept}
Each telescope of the HEGRA IACT system has an azimuthal mount
with 30 mirrors arranged in the so-called Davies-Cotton design~\cite{Davies}.
The total mirror area is A~=~$8.5$~m$^{2}$, 
the mirror dish has a diameter
of D~=~$3.4$~m and a focal length of f~=~$4.90$~m.
The camera is mounted in the primary focus, and consists of
271 PMs in a hexagonal pattern. Its field of view (FOV) is $4.3^\circ$;
each pixel covers $0.25^{\circ}$ opening angle.

The telescope axes are driven by stepper motors; one step turns the
telescope by 1.3''. The axes' positions are monitored by optical
shaft encoders whose resolution is 1.3'. Tracking is computer controled;
the algorithm uses a combination of calculated driving velocity and position
feedback of the shaft encoders to steer the motors. This keeps the telescopes
smoothly on track within $\pm 1$ shaft encoder tick.

A rough ($\approx 0.1^{\circ}$) offline check 
of tracking performance
can be applied if stars of sufficient brightness appear in the FOV
during data taking.
Star images can be reconstructed using the DC currents in the PMs.
However the reconstruction accuracy depends on the relation of
point spread function\footnote{defined as the spot size in the
camera caused by a point light source at infinity, such as a star}
to the pixel size. If the point spread function is small compared to
the pixel size (as it should ideally be the case), the image cannot
be localized to better than the pixel size. Only with a wider point 
spread function, the center of gravity of the currents of several
adjacent pixels provides a better measure. Even in this case problems
arise for irregular point spread functions, as observed near the
edge of the FOV.
For telescopes such as the HEGRA IACTs, with a relatively narrow
point spread function~\cite{Wiedner_optics}, it is therefore inconvenient
to use stars for online tracking control.

With the initial data from the HEGRA IACT system, the need for an
improved calibration of telescope pointing became obvious:
\begin{itemize}
\itemsep0.0cm
\parsep0.0cm
\item Since pointing deviations translate directly into deviations in the 
      reconstructed shower directions - for certain geometries even
      with significant amplification factors - 
      the pointing accuracy should be about one order of magnitude better than
      the air shower reconstruction accuracy which is better than 
      $0.1^{\circ}$~\cite{Crab_system,Mkn501_system}. No significant
      systematic errors should remain, which could shift a point source, or 
      smear out a signal.
\item Pointing corrections should be understood in terms of a model of
      (mainly mechanical)
      error sources. Such a description allows to reliably 
      interpolate the corrections in between a finite set of
      calibration points. One can monitor the time variation
      of the relevant parameters of mount and drive, and may
      ultimately be able to cure problems at their origin.
\end{itemize}

\section{Pointing calibration}
\label{calibration}

Part of the information required for the pointing corrections could
be derived from TV cameras mounted on the mirror dish monitoring
stars as well as bending of the camera masts. The optical path
however is additionally affected by the adjustment of the mirror
segments relative to the dish as well as by deformations of the mirror
dish itself.

Only with star images in the PM camera the full optical path can be
tested. The drawbacks stated above of this way of pointing calibration
can be overcome by dedicated calibration measurements for each star,
so-called {\it point runs}, where
\begin{itemize}
\itemsep0.0cm
\parsep0.0cm
\item the star is focused in the camera center where the spot is regular
      (it can be described by a spherically symmetric 2-dimensional gaussian);
\item the star image is then scanned across the central pixel,
      following a two-dimensional grid with spacings which are
      smaller than both the pixel size and the point spread function.
\end{itemize}

\begin{figure}[h]
\begin{center}
\addtolength{\textwidth}{-2.0cm}
\psfig{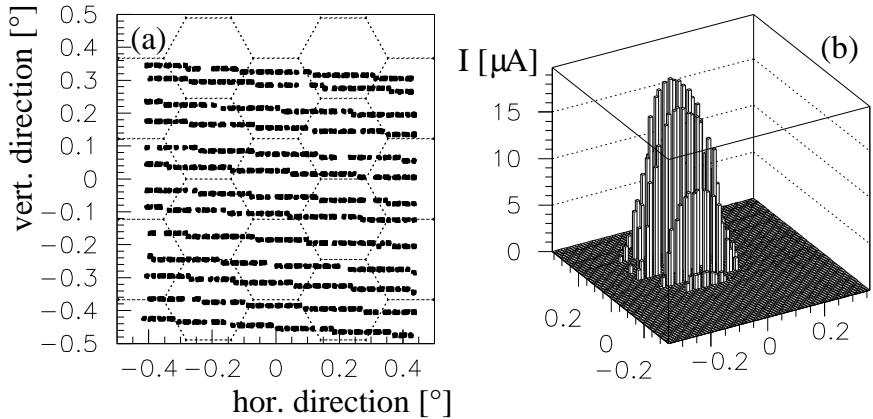}
\caption{Typical example of a point run, where a star is scanned
across the central pixel. Each scan line is obtained by moving the azimuth motor
(horizontal direction), while the altitude motor (vertical direction) is fixed\protect\footnotemark.
The scanlines appear slightly rotated in the camera, because
the star is moving at the same time.
(a) Nominal positions of the star image in the camera, also showing the pixel outlines.
(b) Current in the central pixel, as a function of the distance between the expected star image
and the pixel center;
one notices that the response is not centered at 0, i.e. that there
is a deviation in telescope pointing.}
\label{fig_scanlines}
\end{center}
\end{figure}

\footnotetext{
Horizontal and vertical coordinates refer to a coordinate system fixed to the camera.
With the telescope pointing to the horizon, the horizontal direction is equivalent to azimuth,
and the vertical direction equivalent to altitude.}

\begin{figure}[h]
\begin{center}
\addtolength{\textwidth}{-2.0cm}
\psfig{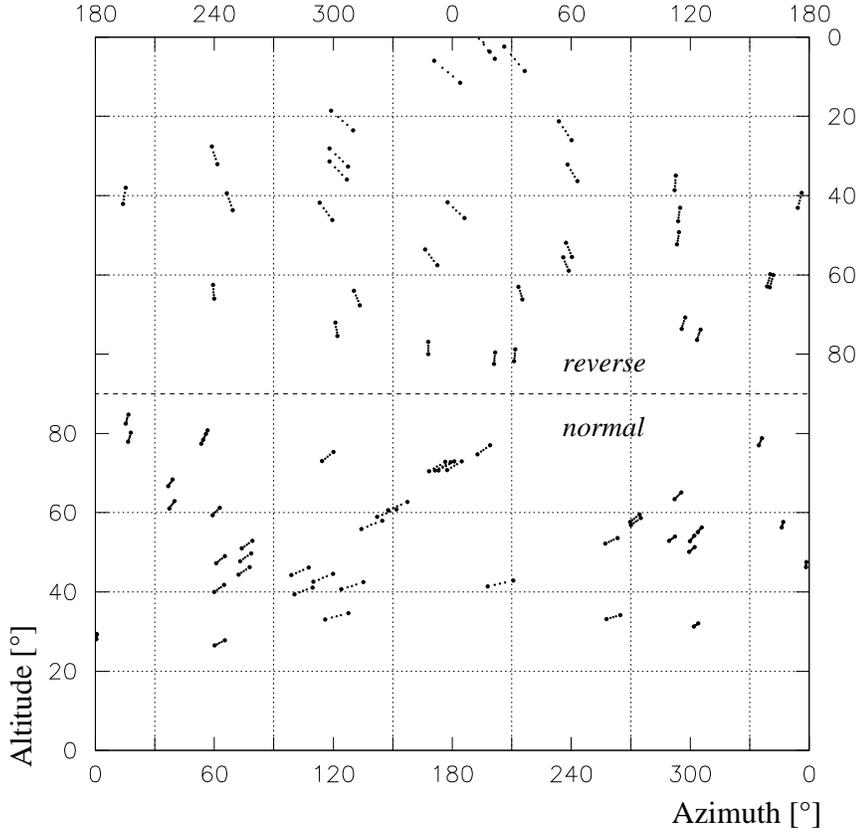}
\caption{Result of a pointing calibration campaign for one telescope.
 The measured
deviations from the expected position are indicated by 
vectors, scaled up by a factor 50. The hemisphere
is doubled due to the two operation modes (see text).}
\label{skyview}
\end{center}
\end{figure}

The current procedure for such point runs represents an evolution of
earlier techniques such as described in~\cite{ct1}.
To minimize the scan time, scans are taken along scan lines with
fixed altitude (Fig.~\ref{fig_scanlines}(a)). The azimuth drive is operated
at constant speed, and PM currents are monitored every time the 
azimuth shaft encoder changes, 
together with the current values of the shaft encoders. A typical
scan comprises 13 such scan lines, with $0.07^{\circ}$ spacing in altitude,
and takes about 15 min.
The resulting distribution of current in the central pixel 
(Fig.~\ref{fig_scanlines}(b)) is proportional to the 
integral of the star's image over the pixel surface. 
The center of the distribution
determines the center of the star image and therefore
the telescope pointing relative to the star. This information 
is then used to correct the pointing determined on the basis of
shaft encoder values. The steepness of the transitions near the
pixel border allows to estimate the width of the point spread function.

For stars of at least 4th (blue) magnitude,
a typical point run procedure provides an accuracy 
of the order of better than $0.01^{\circ}$ per
measurement. To derive a calibration which is valid for the complete hemisphere
a set of pointruns has to be made which covers the whole sky. 
The result of such a
calibration campaign for one telescope is shown in Fig.~\ref{skyview}. 
For each pointrun the deviation
from the expected position is indicated as a vector 
(scaled up by a factor of about 50).
Due to the azimuthal mount the telescope can be driven in two modes
 called {\it normal}
and {\it reverse} (telescope turned across the zenith), 
therefore the hemisphere
must be examined twice.

The minimum number of pointruns is determined by the number of
required parameters to describe the deviations (see section~\ref{model}). 
Experience shows that at least 30 pointruns are needed for
a new calibration which results in a total calibration time
of 8 to 10 hours. For further checks of pointing performance
the dataset may be reduced to a smaller number.

\section{Modeling of pointing deviations}
\label{model}

Various types of misalignments were considered as causes for the measured deviations;
the following effects turned out to be most prominent:
\begin{itemize}
\itemsep0.0cm
\parsep0.0cm
\item shaft encoder non-linearities;
\item zero-point offsets of the shaft encoders;
\item shift of the camera center with respect to the optical axis of the telescope;
\item bending of telescope mount components;
\item non-vertical alignment of the azimuth axis;
\item non-perpendicular alignment of the altitude axis 
relative to the azimuth axis.
\end{itemize}
The influence of some of these parameters can be demonstrated by plotting horizontal
respectively vertical deviation versus either azimuth or altitude while fixing the
other coordinate. In the following two of these four plots
are shown.

\begin{figure}[h]
\begin{center}
\addtolength{\textwidth}{-2.0cm}
\psfig{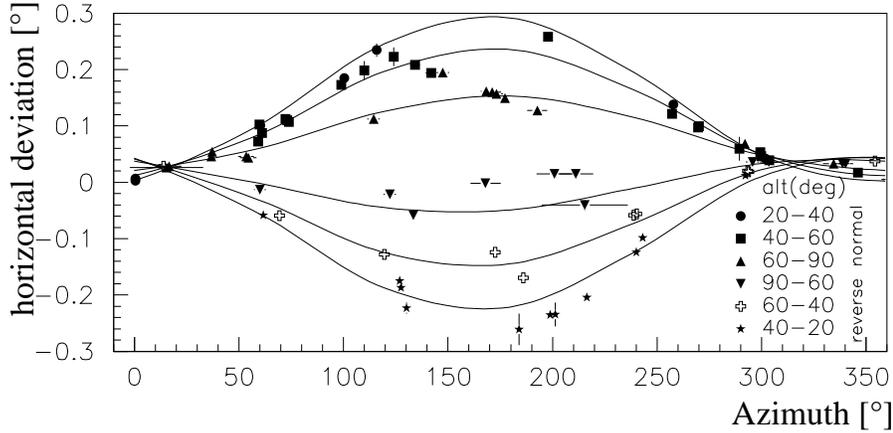}
\caption{Horizontal deviation vs. azimuth while applying six altitude slices.
The main deviation is caused by the azimuth shaft encoder (see text).}
\label{slice1}
\end{center}
\end{figure}

In Fig.~\ref{slice1} the horizontal deviation is plotted versus azimuth. Besides a
small mean offset the main effect here is a harmonic modulation with 
a period of $360^{\circ}$.
This can be interpreted as an error of the azimuth shaft encoder which leads
to a modulation of the azimuth value;
therefore the horizontal deviation decreases
with increasing altitude - at the zenith an error in the azimuth value
is obviously without consequences.

\begin{figure}[h]
\begin{center}
\addtolength{\textwidth}{-2.0cm}
\psfig{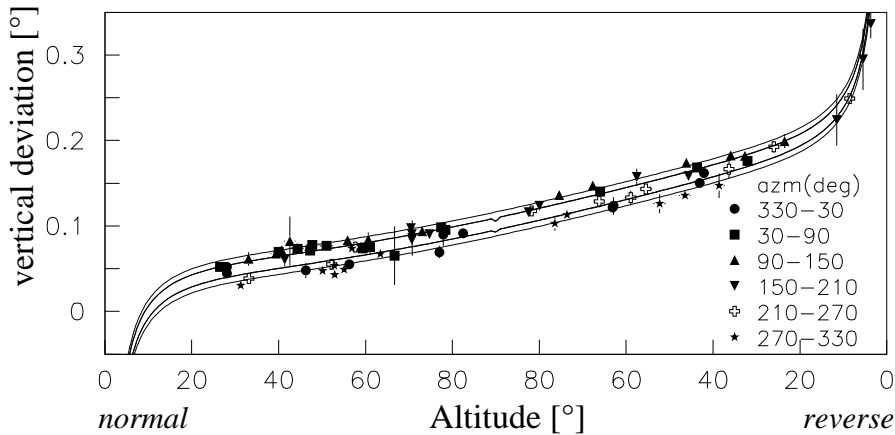}
\caption{Vertical deviation vs. altitude while applying six azimuthal slices.
The common behaviour at all azimuth positions is caused by bending of telescope
mount parts (see text).}
\label{slice2}
\end{center}
\end{figure}

In Fig.~\ref{slice2} the vertical deviation is plotted versus altitude.
For all azimuth slices a common trend is visible which is interpreted as bending
of parts of the telescope mount. 
This leads to a deviation in the vertical direction which
depends on the altitude position. 
The small differences at different azimuth positions are caused by the
non-vertical alignment of the azimuth axis.

At small altitudes atmospheric refraction becomes visible, causing a
steep variation of pointing deviations with altitude.

\begin{figure}[h]
\begin{center}
\addtolength{\textwidth}{-2.0cm}
\psfig{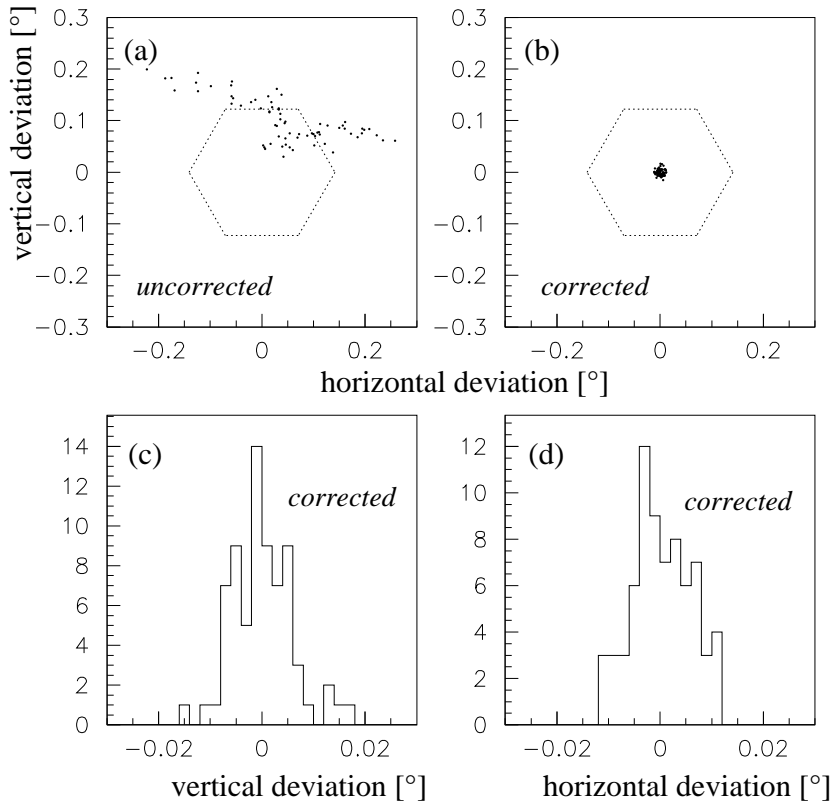}
\caption{The measured deviations are shown in the camera system (a) without correction,
(b) after correction using the model described in the text. 
For reference the outline of the
central pixel is superimposed. (c) and (d) show the remaining deviations in both
coordinates, after correction.}
\label{result}
\end{center}
\end{figure}

All the sources of pointing deviations, including atmospheric refraction,
are summarized in a model where their amplitudes and, where applicable, their
phases are treated as free parameters. The parameters are determined from a
fit to the set of measured pointing deviations. The model can then be used
to interpolate corrections between calibration points. As a measure of the
quality and consistency of the model, Fig.~\ref{result} summarizes the pointing
deviations before (a) and after (b) the correction function is applied. After
corrections, the rms pointing error in both the horizontal (c) and vertical (d)
directions is about $0.005^{\circ}$, with a mean consistent with zero.
We note that the number of fit parameters (11) is significantly smaller
than the number of measured points, each of which provides two coordinates,
($2 \times 71$ in the dataset used here for demonstration,
about $2 \times 30$ in standard new calibrations),
hence this result is by no means trivial, and provides a genuine check of the model.

\section{Results}
\label{results}

\begin{table}[t]
\begin{center}
\small
\begin{tabular} {|l|c|c|c|c|} 
\hline
Telescope        &  CT3 & CT4 & CT5 & CT6 \\
\hline
maximum bending                     & $ 0.039^{\circ}$ & $ 0.068^{\circ}$ & $ 0.086^{\circ}$ & $ 0.077^{\circ}$   \\
vertical camera offset              & $ 0.049^{\circ}$ & $ 0.067^{\circ}$ & $-0.055^{\circ}$ & $ 0.059^{\circ}$   \\
horizontal camera offset            & $ 0.034^{\circ}$ & $ 0.004^{\circ}$ & $ 0.053^{\circ}$ & $-0.018^{\circ}$   \\
azimuth offset                      & $-0.037^{\circ}$ & $ 0.002^{\circ}$ & $-0.109^{\circ}$ & $ 0.095^{\circ}$   \\
non-vertical alignment azimuth axis &                  &                  &                  &                    \\
~~~amount                           & $ 0.014^{\circ}$ & $ 0.039^{\circ}$ & $ 0.041^{\circ}$ & $ 0.016^{\circ}$   \\
~~~phase (sky direction)            & $ 224.8^{\circ}$ & $ 329.3^{\circ}$ & $ 199.1^{\circ}$ & $ 161.4^{\circ}$   \\
non-linearity azimuth shaft encoder &                  &                  &                  &                    \\
~~~amplitude                        & $ 0.172^{\circ}$ & $ 0.024^{\circ}$ & $ 0.030^{\circ}$ & $ 0.028^{\circ}$   \\
~~~phase                            & $ 287.4^{\circ}$ & $ 308.0^{\circ}$ & $  33.8^{\circ}$ & $ 277.2^{\circ}$   \\
non-linearity altitude shaft encoder&                  &                  &                  &                    \\
~~~amplitude                        & $ 0.024^{\circ}$ & $ 0.002^{\circ}$ & $ 0.154^{\circ}$ & $ 0.018^{\circ}$   \\
~~~phase                            & $ 179.2^{\circ}$ & $ 180.0^{\circ}$ & $   1.1^{\circ}$ & $ 180.1^{\circ}$   \\
non-perpendicularity altitude axis  & $ 0.022^{\circ}$ & $ 0.045^{\circ}$ & $ 0.029^{\circ}$ & $ 0.036^{\circ}$   \\
\hline
\end{tabular}
\vspace{5ex}
\normalsize
\caption{Values of all correction parameters for CT3 - CT6 as they were valid in February 1997}
\label{restab}
\end{center}
\end{table}

\begin{table}[b]
\begin{center}
\small
\vspace{2ex}
\begin{tabular} {|l|c|c|c|c|}
\hline
Telescope        &  CT3 & CT4 & CT5 & CT6 \\
\hline
before correction & & & & \\
~~~rms horizontal  & $0.079^{\circ}$ & $0.023^{\circ}$ & $0.085^{\circ}$ & $0.057^{\circ}$ \\
~~~rms vertical    & $0.043^{\circ}$ & $0.068^{\circ}$ & $0.082^{\circ}$ & $0.065^{\circ}$ \\
\hline                     
after correction & & & & \\
~~~rms horizontal  & $0.005^{\circ}$ & $0.006^{\circ}$ & $0.011^{\circ}$ & $0.005^{\circ}$ \\
~~~rms vertical    & $0.005^{\circ}$ & $0.006^{\circ}$ & $0.013^{\circ}$ & $0.008^{\circ}$ \\
\hline                     
\end{tabular}              
\vspace{5ex}
\normalsize                
\caption{Rms pointing deviations measured with typically 30 stars,
in the horizontal and vertical directions, before and after pointing
corrections, for the four HEGRA telescopes operating in stereoscopic
coincidence mode.}
\label{tab_rms}
\end{center}
\end{table}

The correction functions were determined for all four HEGRA telescopes
(CT3, CT4, CT5, CT6) 
participating in stereoscopic mode. 
Table~\ref{restab} shows the correction parameters
which were valid in February '97. Typical errors supplied by the fit are
$0.005^{\circ}$ or less for the
amplitudes and $5^{\circ}$ for the phases, respectively. Provided that the 
calibration points are reasonably spread across the sky, each of the
correction terms has its own distinctive signature, and the 
correlations between fit parameters are modest. Repeated complete
calibrations demonstrated that the correction parameters are reproducable,
and rather stable in time, unless hardware changes intervene (see below).

For all four telescopes, the correction scheme succeeded in reducing 
pointing errors from initial values of order of the pixel size, to
acceptable ranges of the order of $0.01^\circ$ or less, see Table~\ref{tab_rms}.
The telescope CT5 is slightly worse than the others, possibly indicative of an additional
source of pointing deviations, which is not modeled in the fit.

\begin{figure}[b]
\begin{center}
\addtolength{\textwidth}{-2.0cm}
\psfig{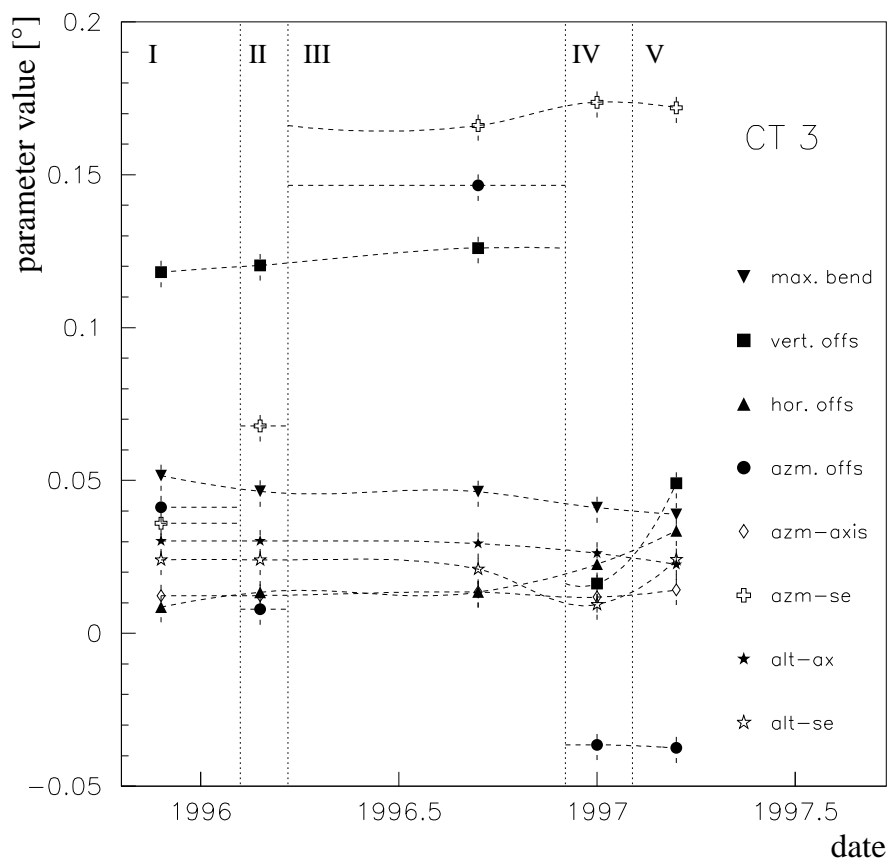}
\caption{Time evolution of the correction parameters of IACT 3 
(only amplitudes).
The dotted vertical lines separate the periods in which one correction function
was valid. The dashed lines are drawn to guide the eye.
Period I/II and II/III: reinstallation of azimuth shaft encoder.
Period III/IV: redefinitions of shaft encoder zero positions. Period IV/V: no
hardware change.}
\label{history}
\end{center}
\end{figure}

The observed pointing deviations before correction, 
and hence the amplitudes of the correction parameters are typically
of the order of the pixel size or less. This implies that tracking is good
enough to maintain a source reasonably centered within the field of view,
and that it is sufficient to apply pointing corrections offline, as first
step of the analysis procedure. 
The great advantage of such a procedure is that after hardware changes
which affect the pointing (see below) the calibration procedure can be delayed
to a convenient time.

The interpretation of the parameters is straight forward. Only
the modulation of the altitude shaft encoder of IACT 3, 4 and 6 
(see Table~\ref{restab})
may also be interpreted as a common change of 
the bending behavior between normal
and reverse mode because of similar amplitudes and phases.

Hardware changes which demand a new calibration are,
for example, readjustments of mirrors or the
exchange of shaft encoders. The effect of such an exchange 
can be seen in Fig.~\ref{history}
where the time evolution of the correction parameters of telescope CT3
is shown, since the beginning
of operation. Due to repair work the azimuth shaft encoder
had to be reinstalled twice (between periods I,II and between II,III).

Without any preceding installation work, so far only changes in the offsets
were observed. These changes are interpreted as slight changes in the mirror
adjustment which lead to a broadening of the point spread function as well
as to a shift of its center (p.e. Period IV/V, vertical offset).

\section{Application to air shower observations}
\label{application}

The pointing calibration was tested with two (within resolution)
point sources of known position:
\begin{itemize}
\itemsep0.0cm
\parsep0.0cm
\item the Crab pulsar, the TeV ``standard candle'', a plerion at a distance of 8.5 kpc;
\item Mkn 501, a blazar at a redshift of $z \approx 0.03$.
\end{itemize}
Analysis procedures and 
first results concerning the Crab observations were reported in
\cite{Crab_system}, still using preliminary and incomplete pointing
corrections. The results on $\gamma$-ray emission from Mkn 501, as 
summarized in \cite{Mkn501_letter}, used essentially the final 
corrections as described in this paper.

\begin{figure}[b]
\begin{center}
%\addtolength{\textwidth}{-2.0cm}
%\psfig{file=sources_02_loosecuts.eps,width=\textwidth}
\addtolength{\textheight}{-3.7cm}
\psfig{file=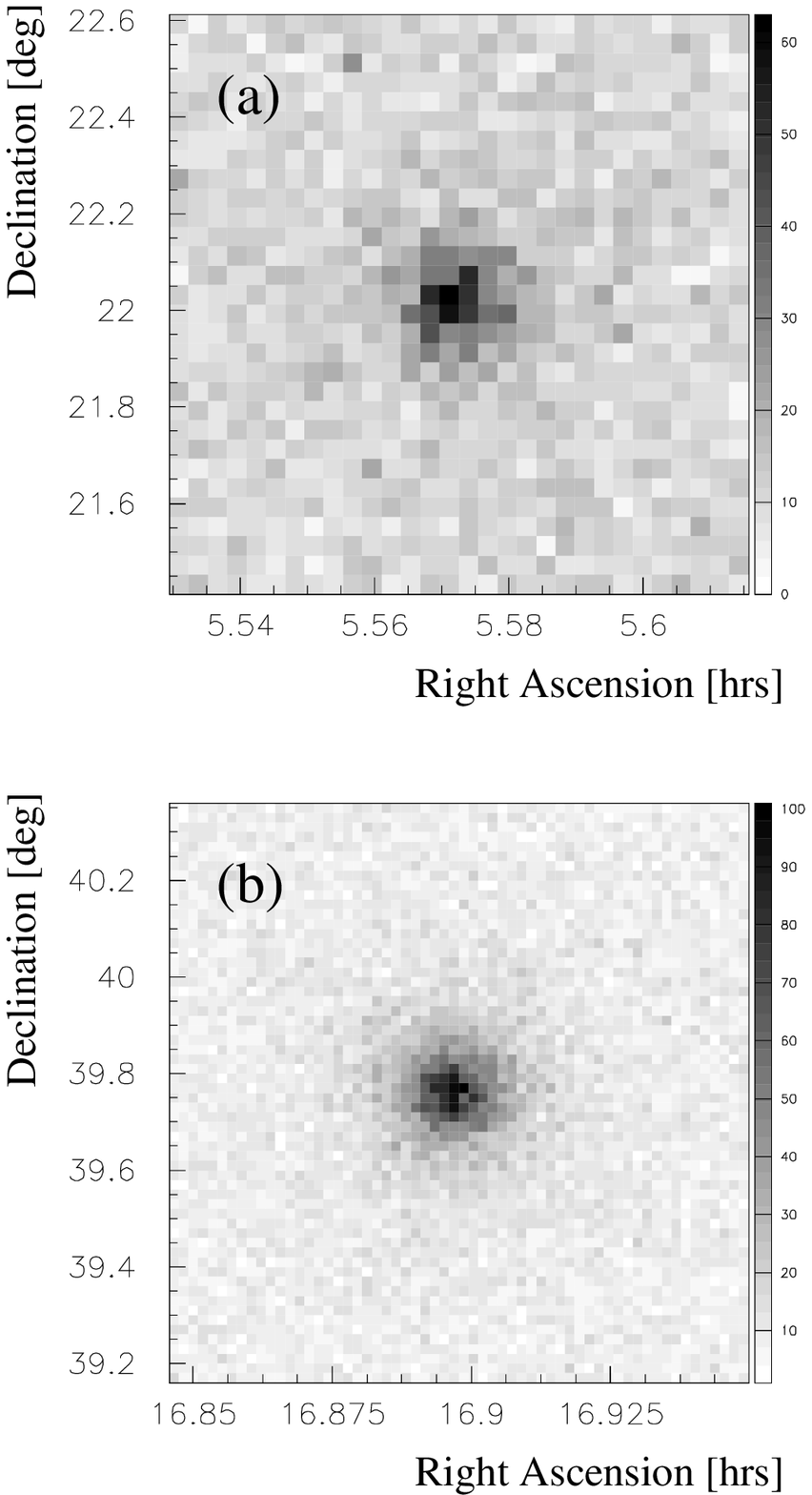,height=\textheight}
\caption{Excess of $\gamma$-ray candidate events from the direction of
the Crab nebula (a) and the blazar Mkn 501 (b).
A cut on the mean width of the images,
averaged over all telescopes, is used to separate $\gamma$-ray candidates
from the uniform cosmic-ray background.}
\vspace{3ex}
\label{fig_sources}
\end{center}
\end{figure}

\begin{table}[h]
\begin{center}
\small
\begin{tabular} {|l|c|c|} 
\hline
Source                           &  Crab              &    Mkn 501 \\ 
\hline
Observation period               &  12/96 - 2/97      &    3/97 - 5/97    \\
On-source time                   &  24.3 hrs          &    62.7 hrs        \\
Observation mode                 &  normal            &    reverse         \\
\hline
Radio source position of current epoch  &             &                   \\
~~~ra                            &    5.5725 hrs      &     16.8963 hrs      \\
~~~dec                           &  $22.013^{\circ}$  &    $39.764^{\circ}$  \\
\hline
$\gamma$-ray source position     &                    &            \\                     
~~~ra                            &    5.5720 hrs      &     16.8968 hrs     \\                     
~~~dec                           &  $22.020^{\circ}$  &    $39.762^{\circ}$  \\
\hline
Angular distance radio source - $\gamma$-ray source   &         &   \\
~~~along ra                      &  $0.007^{\circ} \pm 0.007^{\circ}$   &    $0.006^{\circ} \pm 0.007^{\circ}$   \\
~~~along dec                     &  $0.007^{\circ} \pm 0.007^{\circ}$   &    $0.002^{\circ} \pm 0.007^{\circ}$   \\  
\hline  
(Gaussian) angular resolution    &  $0.09^{\circ}$    &    $0.09^{\circ}$    \\ 
\hline           
\end{tabular}
\vspace{2ex}
\normalsize
\caption{Data sets and results of the analysis of Crab and Mkn 501 data.
The errors given for the angular distances represent our estimate on
systematical errors; the statistical errors are significantly smaller.}
\label{datatable}
\end{center}
\end{table}

Figs.~\ref{fig_sources}(a) and (b) show the distribution of excess events
in celestial coordinates. The positions of the
$\gamma$-ray sources were determined from fits of two-dimensional 
Gaussian distributions to the excess. The relevant parameters
of the data sets and the fit results concerning the source positions
are summarized in Table~\ref{datatable}.

The reconstructed positions differ from the
radio source positions by less than 1 arcminute which is in good agreement
with the expected pointing accuracy. The corresponding systematic errors are estimated
from the residuals of telescope pointing after correction (see Figure~\ref{result} (c) and (d)).
The angular resolution (characterized
here by the Gaussian width of the distribution of excess events, projected
onto one of the axes) is well below $0.1^\circ~$\footnote{In particular
for the high-statistics Mkn 501 data set, the excess 
is not exactly Gaussian, but has a sharper peak and a more pronounced tail.
Since the distribution represents a wide range of $\gamma$-ray energies,
with the angular resolution improving at high energies, one expects
such deviations from an exact Gaussian shape.}.

\section{Summary}
\label{summary}

The HEGRA IACT system provides a spatial resolution of a new quality in
TeV $\gamma$-ray astronomy (and even in $\gamma$-ray astronomy in general).
The origin of an individual $\gamma$-ray can be reconstructed with an accuracy
of better than $0.1^{\circ}$. The position of a point source of sufficient flux can be 
determined within $0.01^{\circ}$. To achieve this
pointing quality a new calibration procedure for IACTs had to be established.
It uses a set of  (typically $\ge 30$) pointing 
calibration measurements distributed over the
complete sky. These determine the parameters of
correction functions to model the mechanical misalignments of the four
telescopes which are included in the HEGRA IACTS so far. The correction 
functions work for the complete range in azimuth and altitude.

The precise localization of $\gamma$-ray sources allows better
identification of counterparts in other wavebands, and may ultimately
allow the precise mapping of extended sources.

\section*{Acknowledgements}

The support of the German Ministry for Research 
and Technology BMBF and of the Spanish Research Council
CYCIT is gratefully acknowledged. We thank the Instituto
de Astrofisica de Canarias for the use of the site and
for providing excellent working conditions.
We thank the other members of the HEGRA CT groups, who participated
in the construction, installation, and operation of the telescopes.
We gratefully acknowledge the technical support staff of Heidelberg,
Kiel, Munich, and Yerevan.


\begin{thebibliography}{999}

\bibitem{IACTs} T.C.~Weekes, Space Science Rev. 75 (1996) 1;
M. F. Cawley and T.C.~Weekes, Experimental Astronomy 6 (1996) 7.
\bibitem{Hillas} A.M.~Hillas, Proc. 19th ICRC, La Jolla, Vol. 3
(1985) 445.
\bibitem{Punch} M.~Punch et al., Nature 358 (1992) 447.
\bibitem{Petry} D.~Petry et al., Astron. Astrophys. 311 (1996), L13.
\bibitem{Akerlof} C.W.~Akerlof et al., Astrophys. J. 377 (1991), L97.
\bibitem{Crab_system} A.~Daum et al., preprint astro-ph/9704098, and
Astropart. Phys., in press.
\bibitem{Mkn501_system} F.~Aharonian et al., Astropart. Phys. 6 (1997) 343;
Astropart. Phys. 6 (1997) 369.
\bibitem{CT_system} F.~Aharonian, 
Proceedings of the Int. Workshop ``Towards a Major Atmospheric 
Cherenkov Detector II'', Calgary, (1993), R.C.~Lamb (Ed.), p. 81.
\bibitem{Davies} J.M.~Davies, E.S.~Cotton, J. Solar Energy Sci.
and Eng. 1 (1957) 16.
\bibitem{Wiedner_optics} A.~Akhperjanian et al., in preparation.
\bibitem{ct1}  R.~Mirzoyan et al., Nucl. Instr. Meth. A351 (1994) 513.
\bibitem{Mkn501_letter} F.~Aharonian et al., Astron. Astrophys. 327 (1997), L5.

\end{thebibliography}
\end{document}